\documentclass[journal]{vgtc}                     

\onlineid{0}

\vgtccategory{Research}

\vgtcpapertype{Evaluation}

\title{\edit{Review and Analysis of} Scientific Paper Embellishments}

\author{%
  \authororcid{Jiayi Hong}{0000-0002-1332-5045},
  \authororcid{Yixuan Wang}{0000-0003-0195-1193},
  \authororcid{Petra Isenberg}{0000-0002-2948-6417}, and 
  \authororcid{Ross Maciejewski}{0000-0001-8803-6355}
}

\authorfooter{
  \item
  	J.\ Hong is with Universit\'e Laval, Canada and Arizona State University, USA.
        E-mail: jiayi.hong@ift.ulaval.ca.
  \item
        Y.\ Wang, and R.\ Maciejewski are with Arizona State University, USA.
        E-mail: {\{ywan1290\,$|$\,rmacieje\}}@asu.edu.
  \item
  	P.\ Isenberg is with Universit\'e Paris-Saclay, CNRS, Inria, France. 
        E-mail: petra.isenberg@inria.fr.
}

\abstract{%
We present a review and \edit{analysis} of \textbf{scientific paper embellishments}---simple visual elements that are deeply integrated \edit{into} the text of scientific publications. \edit{These embellishments are increasingly used in research papers}, which have the potential to enhance textual descriptions, strengthen connections between figures and content, and improve internal textual coherence, \edit{while also carrying the risk of disrupting the reading experience. As} their exact impact is not yet well understood, \redsout{to leverage and scientifically evaluate these potential benefits,} we conducted a systematic review of all visualization papers published between 2019 and 2024 in IEEE VIS, ACM CHI, and EuroVis. From this corpus, we identified \colCount\ papers that used paper embellishments and distilled three key dimensions that characterize \edit{their usage:} purposes (\why), design choices (\how), and locations (\where) of paper embellishments. Our findings provide a structured perspective on the form of current embellishments in scientific writing in the visualization domain and \edit{provide insights into their role in shaping scientific communication.} 
}


\graphicspath{{figs/}{figures/}{pictures/}{images/}{./}} 

\usepackage{tabu}                      
\usepackage{booktabs}                  
\usepackage{lipsum}                    
\usepackage{mwe}                       
\usepackage{flushend}
\usepackage[normalem]{ulem}
\usepackage{xcolor}

\usepackage{amsthm} 
\usepackage[framemethod=tikz]{mdframed}
\usepackage{tikz}

\definecolor{styledColor}{HTML}{1869a0}

\newtheoremstyle{defi} 
  {\topsep}%
  {\topsep}%
  {\normalfont}%
  {}%
  {\itshape}%
  {:}%
  {.5em}%
  {\thmname{#1}\thmnote{~(#3)}}%
\theoremstyle{defi}
\newmdtheoremenv{definitioni}{Definition}
\newmdtheoremenv[
backgroundcolor=styledColor!10,
hidealllines=true,
leftline=true,
innertopmargin=-4pt,
innerbottommargin=2pt,
linewidth=4pt,
linecolor=styledColor!40,
innerrightmargin=0pt,
]{definitionii}{\it Scientific Paper Embellishments}
\newmdtheoremenv[
roundcorner=5pt,
innertopmargin=0pt,
innerbottommargin=5pt,
linewidth=4pt,
linecolor=styledColor!40,
]{definitioniii}{Definition}

\usepackage{mathptmx}                  
\usepackage[resetlabels,labeled]{multibib}
\newcites{PE}{Full Collection}

\newcommand{\inlinevis}[3]{\raisebox{#1}[0pt][0pt]{\includegraphics[height=#2]{#3}}}

\newlength{\HeightReference}
\newlength{\Width}%
\newcommand{\BGColorBox}[2][styledColor!20]%
{\settowidth{\Width}{#2}%
    \setlength{\fboxsep}{1.5pt}%
    \colorbox{#1}%
    {\raisebox{-0.5mm}%
        {\parbox[b][\HeightReference][c]{\Width}{\centering#2}%
        }%
    }%
}

\newcommand{\dimension}[1]{\textit{#1}}
\newcommand{\background}[1]{\BGColorBox{\textcolor{black}{#1}}}
\newcommand{\colorFont}[1]{\textcolor{styledColor}{#1}}
\newcommand{\outline}[1]{\fcolorbox{styledColor}{white}{#1}}

\newcommand{\eg}{e.\,g.}
\newcommand{\ie}{i.\,e.}
\newcommand{\colCount}{374}

\newcommand{\edit}[1]{\textcolor{black}{#1}}

\newcommand\redsout{\bgroup\markoverwith{\textcolor{red}{\rule[0.5ex]{2pt}{0.4pt}}}\ULon}
\renewcommand{\redsout}[1]{}

\definecolor{whyColor}{HTML}{e8827c}
\newcommand{\why}{\textcolor{whyColor}{\textbf{WHY}}\xspace}
\definecolor{howColor}{HTML}{d1a63d}
\newcommand{\how}{\textcolor{howColor}{\textbf{HOW}}\xspace}
\definecolor{whereColor}{HTML}{8aada7}
\newcommand{\where}{\textcolor{whereColor}{\textbf{WHERE}}\xspace}

\newcommand{\wordScaleGraphics}{\inlinevis{-1pt}{1em}{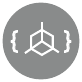}}
\newcommand{\nonDataDriven}{\inlinevis{-1pt}{1em}{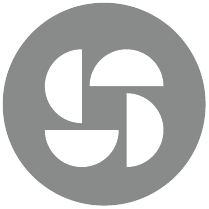}}
\newcommand{\dataDriven}{\inlinevis{-1pt}{1em}{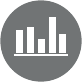}}
\newcommand{\styledPara}{\inlinevis{-1pt}{1em}{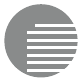}}
\newcommand{\styledWords}{\inlinevis{-1pt}{1em}{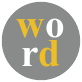}}

\newcommand{\icon}[1]{\inlinevis{-1.5pt}{1em}{#1}}

\usepackage{inputenc}
\makeatletter
\renewcommand\paragraph{\@startsection{paragraph}{4}{\z@}%
  {-0.8ex\@plus -1ex \@minus -.2ex}%
  {-1em}%
  {\normalfont\normalsize\bfseries}}
\makeatother

\usepackage{wheelchart}
\newlength{\myLength}	
\newlength{\mytextsize}
\newcommand{\minipie}[2]{
\raisebox{2pt}[0pt][0pt]{
\resizebox{\myLength}{!}{
\begin{tikzpicture}[baseline = .15\mytextsize]
\wheelchart[
  data sep=0,
  inner data =,
  pie,
  expand list=true,
  contour = #2,
  slices style={
  \WCvarB,
  }
  ,
]{%
  #1/#2/,%
  100-#1/white/
}
 \end{tikzpicture}
}
}}

\teaser{
  \centering
  \includegraphics[width=\linewidth]{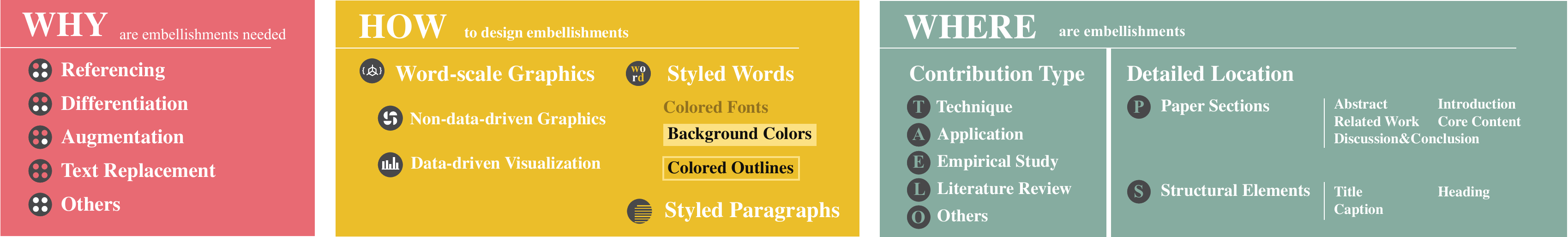}
  \caption{Our proposed \edit{characterizations} for \edit{current practices of} paper embellishments.
  The \edit{characterizations} consist of three dimensions: \why are embellishments needed in scientific papers, \how to design embellishments, and \where are embellishments being placed. 
  }
  \label{fig:teaser}
}

\begin{document}


\firstsection{Introduction}
\maketitle
\label{sec:intro}
Scientific papers are one of the primary means of disseminating research findings. Yet, scientific content is often inherently complex and difficult for researchers to communicate due to complex terminology, dense writing, and the requirements to track and memorize terms and references across the text~\cite{Rowan.1991.SLF}. The European Association for Science Editors has published guidelines~\cite{EAS.2018.EAS} that provide instructions on how to achieve complete, concise, and clear writing. Beyond such dedicated writing and structural guidelines, various other strategies can help readers extract and retain essential information. These strategies include visual aids and graphics~\cite{Victor2011explorable, Rolandi.2011.BGD}, as well as systems designed to support effective reading~\cite{Ponsard.2016.PQV}.

Highlighting textual content to better deliver key points has long been possible using simple approaches via \textbf{bold} or \textit{italicized} text or choosing specific \texttt{font types} and {\large sizes}~\cite{Embley:1982:CWE}. \edit{With text-editing tools} readily available to integrate visual embellishments, such as \colorFont{colored fonts},\background{background highlights}, customized icons~\icon{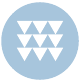}, and even word-scale visualizations~\icon{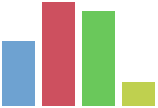}, scientific papers can now incorporate richer, more visually engaging elements. However, the role of such embellishments in scientific papers remains less understood. Key questions arise: How \edit{were and} could embellishments be used in scientific papers? \redsout{Do they enhance scientific communication? And how should authors strategically employ them to improve readability and accessibility?}\edit{How can they potentially impact readability and accessibility?} \redsout{To address such questions, we first need to systematically review current embellishment practices in scientific papers before being able to test their benefits empirically;}These needs motivate our work.

In this manuscript, we focus on the use of what we call scientific paper embellishments, or ``paper embellishments'' for short:
\vspace{-2ex}
\begin{definitionii}
\label{def}
simple visual components that \redsout{seamlessly} \edit{deeply} integrated into the \edit{main} text of scientific publications.
\end{definitionii}
\vspace{-1ex}
The \colorFont{blue} bar on the side of the definition, together with the colored background, is an example of such an embellishment. The \edit{deep} integration of paper embellishments requires close visual and structural alignment of the visual components with the surrounding text. This integration ideally ensures a smooth reading experience without disrupting the reading flow~\cite{Goffin:2017:ESW}.
While some embellishments may carry data, they are often non-essential. Non-essential embellishments can be replaced by plain text, and their absence does not affect comprehension or reading flow. \redsout{Although traditional formatting techniques such as \textbf{bolding} could technically fall within this definition, our focus is not on these traditional typographic tools with limited perceptual novelty.} Many conference templates make use of predefined formatting rules, \edit{such as \textbf{bolding},} suggesting they are already standardized and widely adopted. 
\redsout{Instead, we focus on visual features that add data and/or have artistic and design-oriented features that allow authors to be expressive beyond well-documented functionally conventional elements.}



\edit{On the one hand, the growing number of publications with paper embellishments suggests their potential functions allowing authors to be expressive,} \redsout{Paper embellishments can serve multiple potential functions,} such as emphasizing key points (\eg, the embellishments used in the definition above), establishing references across different sections within a paper, and augmenting textual information. \edit{On the other hand, the lack of clear instructions on how to effectively use these embellishments raises the risk of misuse, which may hinder rather than improve the reading experience. To investigate these aspects, we first need to develop a deep understanding of the current practices, specifically within the context of scientific writing. By reviewing and analyzing these practices, we can identify both the opportunities and challenges associated with paper embellishments, laying the groundwork for future empirical studies to evaluate their specific impacts.}
\redsout{Authors of scientific papers lack a catalog of options for \why paper embellishments may be used, \how to design them, and \where to place them. Once these purposes, design choices, and placements are well-established and understood, the effectiveness of using paper embellishments, as well as their potential benefits and detriments, can be better evaluated via empirical studies.}
%

\redsout{To support authors in designing and employing paper embellishments and guiding future studies on the use of paper embellishments, specifically in our domain,}\edit{To this end,} we conducted a review of the use of paper embellishments in visualization papers. \edit{As a predominantly visual and graphics-oriented domain, we considered recent visualization papers a prime candidate for starting our investigation into paper embellishments.} 
We \redsout{conducted a systematic} reviewed\redsout{of} papers published in IEEE VIS, ACM CHI, and EuroVis from 2019 to 2024 and collected \colCount\ publications that contained at least one paper embellishment.
We categorized all identified paper embellishments along three key dimensions: \why the embellishments were used---whether to reference relevant material, differentiate content, augment text understanding, or replace text; \how they were designed---their visual form; and \where they appeared---with respect to both the contribution type of the paper and the specific sections in which they were embedded.
\edit{Building on this review and analysis, we derived current opportunities and challenges of using embellishments in scientific papers.}

In summary, we offer the following contributions:
\begin{itemize}
    \item A formal definition of paper embellishments as an authoring technique for scientific writing;
    \item \edit{A comprehensive review of current practice of paper embellishments in visualization research;}
    \item \edit{An analysis of the opportunities and challenges in using paper embellishments;}
    \item A research agenda outlining key directions for future investigation on how to effectively leverage paper embellishments in scientific publications.
\end{itemize}

\section{\edit{Background}}
\edit{Scientific articles are a primary medium for scholarly communication. A key aspect of effective scientific writing is readability, commonly defined as ``how easily written materials can be read and understood~\cite{richards:2013:LDL}.'' Researchers have been exploring approaches to improving readability, including recent efforts that incorporate visualizations to facilitate reading~\cite{Oelke.2010.VRA, Hohman.AMN}, develop interactive systems to improve navigation and linkage~\cite{Head.2021.ASP, Wu.2023.FFL, Tao.2025.FFF, Fitzsimmons.2020.ISR}, and employing natural language processing (NLP) models to predict and assess readability with greater precision~\cite{Crossley.2019.MBC}.}

\edit{Embellishments in visualizations include both elements within the visualization itself~\cite{Alebri:2024:ERP, Park:2018:GAP}, such as annotations on charts~\cite{Latif:2021:DUV}, and elements adjacent to visualization, like legends~\cite{Dykes.2010.RML}. 
Similarly, paper embellishments may appear within the main text or be directly applied to the text~\cite{Choudhry.2021.OUT, Sulir:2018:VAS}. 
For instance, prior work has investigated word-scale embellishments that embed data representations within Wikipedia articles~\cite{Goffin:2017:ESW}. 
Other studies have examined the actual use of textual embellishments, \eg, bolding and underlining, to convey emphasis or structure~\cite{Leow:2017:EIP, Han.2008.TEI}.}

\begin{figure}
    \centering
    \includegraphics[width=0.8\linewidth]{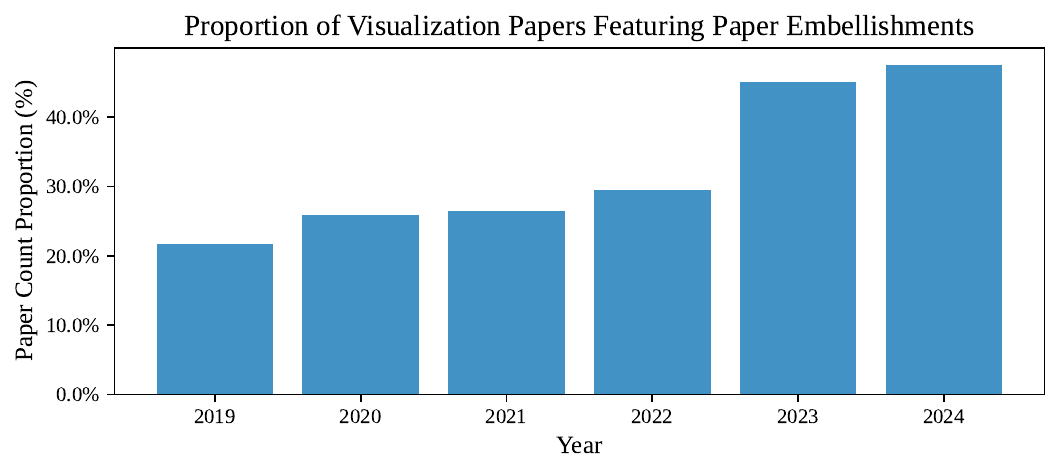}
    \caption{\edit{The trend in the proportion of publications incorporating paper embellishments. The overall percentage of papers using embellishments across IEEE VIS, EuroVis, and ACM CHI shows a clear increase from 2019 to 2024.}
    }
    \vspace{-1em}
    \label{fig:pub_with_pe}
\end{figure}

\edit{The current trend, as shown in \autoref{fig:pub_with_pe}, indicates a stable increase in the proportion of visualization papers using embellishments. In 2019, about $\sim$\minipie{18}{black!50}{\small 18\%} visualization papers across these conferences included embellishment(s), while in 2024, the proportion increased to $\sim$\minipie{41}{black!50}{\small 41\%}. One possible explanation is the growing accessibility of tools that support the integration of such elements. Over the last decades, document editing tools have evolved to incorporate a growing ecosystem of formatting options, including libraries, packages, and macros for adding visual embellishments. 
A second potential reason is the increasing need to strengthen the connection between figures and texts. Scientific papers often present complicated data via visualizations, while readers need to frequently switch between interpreting the figure and understanding the narrative. To maintain the visual appeal while effectively coney contextual relationships, researchers have been exploring techniques to visually link textual descriptions with figures. For example, Steinberger et al.~\cite{Steinberger.2011.CPV} introduced the approach of adding visual links between texts and images, and McColeman et al.~\cite{McColeman.2022.RRV} used leading lines to visually connect these two components.}

\redsout{Effective scientific communication requires carefully chosen methods to enhance public awareness, engagement, interest, and understanding of science~\cite{Burns:2003:SCC}. This includes not only the way information is written but also how it is visually presented and interconnected~\cite{Stokes.2025.IGI}. Authors need to ensure that readers can follow complex descriptions. One strategy authors can employ is to emphasize key elements within a text or incorporate small, in-text illustrations to improve clarity and comprehension~\cite{Goffin:2017:ESW}. To gain a deeper understanding of paper embellishments, it is essential to first examine the use of visual embellishments across various contexts before choosing classes and instances of embellishments for empirical studies on their potential impact on readers.}

\redsout{The visualization community has studied design techniques Tufte called ``chart junk''~\cite{Tufte.1983.VDQ} but which are otherwise called ``visual embellishments.''~\cite{Alebri:2024:ERP, Bateman:2010:UJE} Visual embellishments include icons, images, or graphics added to a visualization that are ``not essential to understanding the data''~\cite{Bateman:2010:UJE}.
Researchers~\cite{Park:2018:GAP} found that adding aesthetic backgrounds to line charts neither reduced error rates nor improved task completion speed. Yet, some research suggests that certain forms of embellishment can be beneficial. For example, Ji et al.~\cite{Ji.2025.TEV} demonstrated that a moderate use of color and bar encodings can help users quickly identify key data points, such as maximum values. Additionally, Andry et al.~\cite{Andry.2021.IEE} analyzed embellishments in infographics and found that they can enhance user engagement and encourage deeper interaction with the content.}

\redsout{Beyond visualization design, researchers have also explored the use of word-scale visualizations embedded within text. Beck et al.~\cite{Beck.2017.WSG} conducted a systematic survey of publications using word-scale visualization across fields, motivated by the potential for these visualizations to avoid requiring readers to cognitively shift between text and separate figures, thereby reducing distractions. However, Goffin et al.~\cite{Goffin.2014.EPD} found that while in-line visualizations can improve accessibility, they may also disrupt sentence flow and make the text harder to interpret in some cases. These works aimed to explore small graphics, but they did not explore the broader space of other embellishments on scientific papers. Specifically, they did not address how word-scale visualizations relate to other types of embellishments, such as annotated text, which can also potentially enhance reading experiences. Beyond small in-line graphics or text annotations, Choudhry et al.~\cite{Choudhry.2021.OUT} also introduced the use of textual narrative techniques (\eg, font size, color, and word-scale graphics) in causality visualization, which could create a more effective reading experience. These works highlight the diverse role of embellishments in scientific communication, offering their potential depending on how they are designed and integrated. Embellishments could function, like visualizations, to better illustrate ideas and provide organized information. In addition, they could also better link figures with descriptive texts. Systematic research such as ours can help to structure options for paper embellishments and aid both designers and researchers in creating and evaluating their potential.}

\section{\edit{Current Practices}}
\label{sec:designspace}
Our work builds upon previous findings by developing \edit{characterizations} for embellishments in scientific papers. 
While recent work in the visualization community has explored the use of word-scale visualizations in textual documents~\cite{Beck.2017.WSG, Goffin.2014.EPD, Goffin:2017:ESW, Goffin.2020.ITV}, word-scale visualizations represent only a subset of paper embellishments commonly found in scientific papers. More recently, a wider range of paper embellishments, such as styled words and customized icons, has become common \edit{(see \autoref{fig:pub_with_pe})}. Our goal is to provide a systematic review of the broad scope of current paper embellishments and to inspire future design innovations.
In this section, we first outline our methodology before introducing and discussing our proposed \edit{characterizations}.

\subsection{Selection Criteria}
\label{sec:criteria}
As we defined in the introduction, paper embellishments are simple visual components integrated into scientific papers. We focused on analyzing embellishments within scientific manuscripts \edit{without} considering appendices or supplementary materials.
We found that many scientific paper templates already predefine formatting rules that automatically apply embellishments, such as adjusting \textbf{font weight} and \textit{shape}. These techniques are frequently used across nearly all scientific research papers to highlight key terms, such as section headers. In our review and collection process, we exclude built-in styling and embellishment techniques that are pre-applied in templates and offer limited perceptual novelty. By examining full paper templates for IEEE VIS, ACM CHI, and EuroVis from 2019 to 2024, we excluded the following embellishment types from our review:
\begin{itemize}
    \item Typographic Techniques: \textbf{weights} (shown in VIS, CHI, EuroVis), \textit{shape} (VIS, CHI, EuroVis), \texttt{family} (VIS, CHI, EuroVis), LETTER CASE (VIS, CHI, EuroVis), {\small font size} (VIS, CHI), black outline (CHI)
    \item Structural Techniques: \href{sec:criteria}{hyperlinks} (VIS, CHI, EuroVis), indentation (CHI, EuroVis), code/equation blocks (VIS, CHI, EuroVis), black inline code/equations (VIS, CHI, EuroVis), footnote (VIS, CHI), numbered lists (CHI)
    \item Symbols: $\bullet$ black bullet symbols (VIS, CHI, EuroVis), mathematical symbols (VIS, CHI, EuroVis), logical operators (CHI), ORCID~\inlinevis{-2pt}{1em}{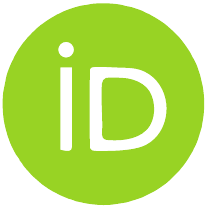} (EuroVis), currency symbols (CHI) 
\end{itemize}
Since the default style of scientific research papers we investigated typically consists of a white background with black text, we also do not consider any of the following as paper embellishments---white background, black text, or black outlines.
%
Some examples, such as leading lines~\cite{McColeman.2022.RRV} and leading backgrounds~\cite{Yang.2024.SPI}, falls into a gray area and are challenging to analyze (see \autoref{app:exclusion}). We excluded these complex structural connections between text and figures, as they often extend beyond the text and link content across different columns.

\edit{During our review process, we also observed that minifigures were used for embellishing textual content~\cite{He.2024.EGA, Yang.2024.SPI}, while we did not focus on them as paper embellishments in our collection due to their ambiguous structural role within the paper layout.
Minifigures sit somewhere between full figures and inline paper embellishments. Unlike most paper embellishments, minifigures span multiple line heights and visually break the text flow, though they typically cover only a part of a column width. At the same time, minifigures lack the formal properties of full figures. They do not have figure numbers, captions, or explicit references in the main text, which makes them less formalized in their communicative function.
In the future, it would be interesting to evaluate the efficacy of minifigures to provide valuable insights into their role as a middle ground between full-size figures and word-scale graphics. By balancing the need for clear compact visual information with the goal of reducing disruptions to reading flow, minifigures may offer a promising embellishment for scientific communication.}

\subsection{Method}
We reviewed 1470 full papers from IEEE VIS, EuroVis, and ACM CHI (with the keyword ``visualization'' in the title or abstract) published between 2019 and 2024 from their respective digital libraries. One author systematically skimmed through the entire set of papers, identifying papers containing embellishments. The author then filtered out papers that did not meet our selection criteria (\autoref{sec:criteria}) and took notes on relevant findings, such as the screenshot of the design, location, and potential function of the embellishment. All the authors then collaboratively conducted open coding, which led to the development of an initial set of design dimensions.
Following this, the same author who reviewed the papers assigned each paper to the identified codes, continuously noting potential refinements needed to better capture the full range of embellishments. After the first round of coding, all authors reconvened to evaluate the coding structure, ensuring clarity and coherence.
To further validate our code set, the same author selected ten papers with dense paper embellishments, which were open-coded collaboratively with a second author. We also developed and refined a codebook throughout this process to ensure consistency in coding. Following this, the second author independently reviewed each paper to verify its categorization. Any disagreements were resolved through group discussions until a consensus was reached, and the codebook was updated accordingly. Given these updates, we conducted an additional round of coding. As a result, each paper was examined by at least two authors at least three times, ensuring the accuracy of our coding.

In total, we identified 12,149 paper embellishments across \colCount\ papers in our collection (243 IEEE VIS papers, 67 CHI papers, and 64 EuroVis). 
\edit{We organized our qualitative coding according to the following dimensions:}\\
\why they are added---examines the function of using paper embellishments in scientific papers (\autoref{Sec:WHY});\\
\how they are designed---explores different types of paper embellishment designs (\autoref{sec:HOW}); and \\
\where they appear in the paper---investigates the contribution types of papers that incorporate embellishments and their specific locations within the manuscript (\autoref{sec:WHERE}). 

\noindent We will use the icons we designed for each code of \how throughout the manuscript (\icon{icons/word_scale_graphics.pdf} \icon{icons/non_data_driven.pdf} \icon{icons/data_driven.pdf} \icon{icons/styled_words.pdf} \icon{icons/styled_para.pdf}), aiming to enhance clarity.

\subsection{\why: Functions of Paper Embellishments}
\label{Sec:WHY}
To understand why scientific papers use paper embellishments, we identified four distinct functions: \dimension{Referencing}, \dimension{Differentiation}, \dimension{Augmentation}, and \dimension{Text Replacement}. 
Initially, our function codes also included three additional purposes: \textit{Emphasis}, \textit{De-emphasis}, and \textit{Aesthetics}. Emphasis is intended to make readers spontaneously differentiate something from surrounding, differently colored (mostly black) text, while De-emphasis aims to make content spontaneously blend into the surrounding texts. However, through continuous discussion, we decided to exclude these two categories. The primary reasons for this exclusion were: 1. Pervasiveness---nearly every embellishment inherently serves the function of either emphasis or de-emphasis, making them redundant as distinct categories; 2. Subjectivity---determining whether an embellishment primarily serves as emphasis or de-emphasis is highly interpretative and inconsistent across different contexts. For example, authors used \textcolor{gray}{gray} text (which could also be seen as black with transparency) to present their hypotheses~\citePE{Araya.2020.ACG}. This choice could be interpreted in two ways: as a form of emphasis, distinguishing the text from standard black text, or as de-emphasis, making it appear closer to the white background. Similarly, we found it difficult to categorize \textit{Aesthetics} as a distinct purpose. Almost all embellishments are designed to enhance visual appeal, but determining whether aesthetics is their purpose is inherently subjective. Therefore, we refined our categorization to focus on four core functions that more clearly differentiate the functions of embellishments.

When coding, we focused on identifying all the primary function(s) of each embellishment. Since the same type of embellishment could appear multiple times within a paper--sometimes with varying functions--we considered all of its observed functions when summarizing. For instance, previous work used numbers (1--6) with varied background colors to encode six pattern types~\citePE{Lan.2021.UNL}. While some numbers appeared with corresponding pattern names and others did not, we consistently classified all such elements of that paper with the function \dimension{Text Replacement}, as they can replace the pattern names. We also excluded cases of syntax highlighting (\eg, \inlinevis{-1.5pt}{1em}{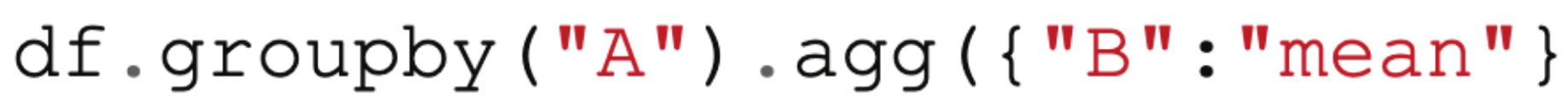}) embedded in the text~\citePE{Epperson.2022.LAH}. In these cases, no reference was made to the core manuscript text, and we did not categorize examples such as this as \dimension{Referencing}. Similarly, some papers included quotes from participants and added embellishments to them, like the example in \autoref{fig:examples} \icon{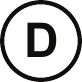}, implicitly referencing their collected qualitative data. We did not code them with \dimension{Referencing}~\citePE{Kruchten.2024.MBE, Kauer.2021.TPL} as the exact wording never appeared elsewhere in the manuscript. 

\subsubsection{Referencing}
\begin{figure*}[t]
    \centering
    \includegraphics[width=\linewidth]{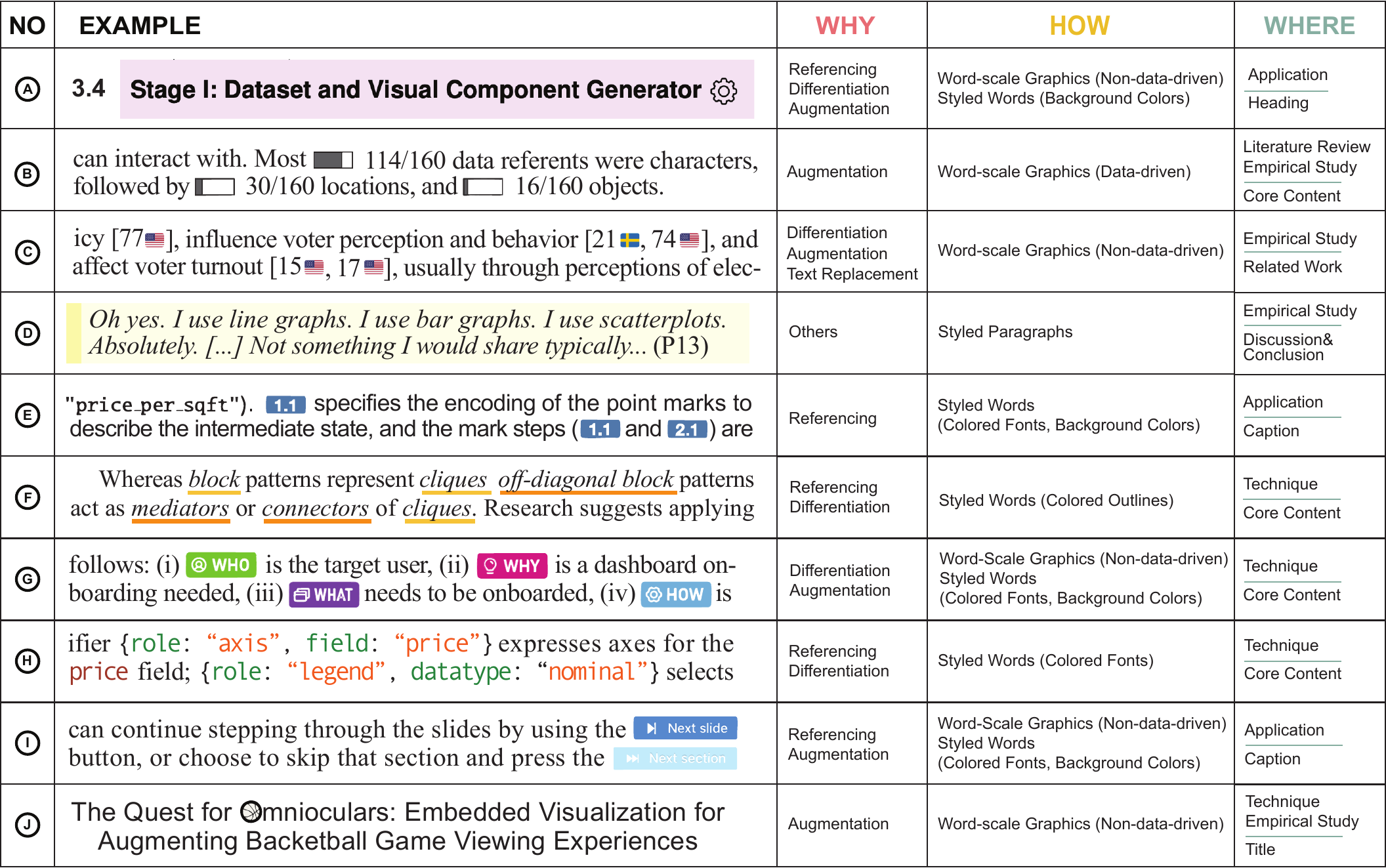}
    \caption{Examples of representative paper embellishments we collected from the literature, covering all categories in our proposed \edit{characterizations}: \icon{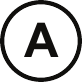} \cite{Cui.2025.PPU} Authors incorporated \dimension{Styled Words} (\dimension{Background Colors} and \dimension{Colored Fonts}) and \dimension{Non-data-driven Graphics} in the Section \dimension{Heading} of an \dimension{Application} paper. These embellishments served multiple functions: \dimension{Differentiating} stages using colors, providing a visual \dimension{Reference} for subsequent mentions of stage I, and augmenting descriptions to enhance understanding of the detailed stages. \icon{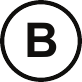} \cite{Yao.2025.UEV} Authors employed \dimension{Data-driven Visualizations} to represent proportions in \dimension{Core Content} for \dimension{Augmenting} the scientific content. \icon{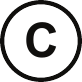} \cite{Yang.2024.SPI} Authors used \dimension{Non-data-driven graphics} (nation flags) in \dimension{Related work} to achieve the functions of \dimension{Differentiate, Augmentation, and Text Replacement}. \icon{icons/d.pdf} \cite{Newburger.2024.VAS} Authors used \dimension{Styled Paragraphs} to \dimension{Differentiate} texts from the plain black text in the \dimension{Discussion and Conclusion} of an \dimension{Empirical Study}. \icon{icons/e.pdf} \cite{Kim.2021.GGR} Authors incorporated \dimension{Background Colors} in the \dimension{Caption} of an \dimension{Application} paper to enhance \dimension{Referencing} to the figure. \icon{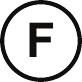} \cite{Fuchs.2025.QMR} Authors utilized \dimension{Colored Outlines} in the \dimension{Core Content} of a \dimension{Technique} paper to support \dimension{Differentiation and Referencing}. \icon{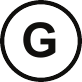} \cite{Dhanoa.2022.PMD} Authors combined \dimension{Word-scale Graphics} and \dimension{Styled Words} to \dimension{Differentiate} dimensions in a \dimension{Technique} paper. \icon{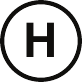} \cite{Kim.2022.CDG} Authors used \dimension{Styled Words} for inserting codes in texts for \dimension{Referencing} detailed codes and \dimension{Differentiating} diverse variables. \icon{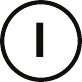} \cite{Li.2023.NND} Authors used the same buttons from the interface which combined \dimension{Word-scale Graphics} and \dimension{Styled Words} for \dimension{Referencing and Augmentation}. \icon{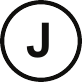} \cite{Lin.2023.QOE} Authors \dimension{Augmented} the paper Title using \dimension{Word-scale Graphics}.}
    \label{fig:examples}
\end{figure*}

These paper embellishments were designed to guide readers to specific content within the manuscript, such as particular parts of figures, tables, key sentences, or algorithms. Authors typically copied the embellished referents, embellished texts, or graphics, and incorporated them into the text to strengthen their connections.
For example, when we point to \autoref{fig:examples} \icon{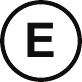} in our own paper we use the icon \icon{icons/e.pdf} to refer to a specific position in the figure. 
From our analysis, \dimension{Referencing} was the most common function we found. We observed that 325 papers used paper 10,395 embellishments for \dimension{Referencing}, accounting for around\minipie{42}{black!50}{\small 42\%} of all the embellishments.

\subsubsection{Differentiation} 
\dimension{Differentiation} occurs when paper embellishments are used to make elements visually distinct from each other within the larger group. Taking our manuscript as an example, we used three colors to \dimension{Differentiate} our three dimensions. Differentiation elements are interdependent and collectively contribute to the overall group, similar to how dimensions in a design space or conditions in an empirical study work together within their respective frameworks. Authors typically applied the same type of embellishments to each element within the group while using diverse design variations (\eg, color, shape, and texture). For example, as illustrated in \autoref{fig:examples} \icon{icons/g.pdf}, the authors embedded different combinations of graphics and background colors on dimensions to \dimension{Differentiate} dashboard onboarding space~\cite{Dhanoa.2022.PMD}.

In total, there were 301 papers that used paper embellishments for \dimension{Differentation} with 9,748 embellishments in total. This indicates that at least\minipie{39}{black!50}{\small 39\%} of all embellishments served this function, though some may have had additional functions.

\subsubsection{Augmentation}
This function includes paper embellishments which are employed for clarification, simplifying complex content, and adding supplementary information. In \edit{the} Introduction, we used graphics \icon{icons/customized_icon.pdf} \icon{icons/word_scale_vis.pdf} to enhance the clarity and meaning of the words `icons' and `word-scale visualizations'. Similarly, authors illustrated texts with emojis~\icon{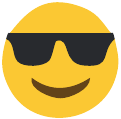} \cite{Wang.2023.GAM}, reinforced descriptions through graphics (Screenfit~\icon{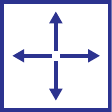})~\cite{Bach.2023.DDP}, or conveyed additional details using data-driven visuals (\icon{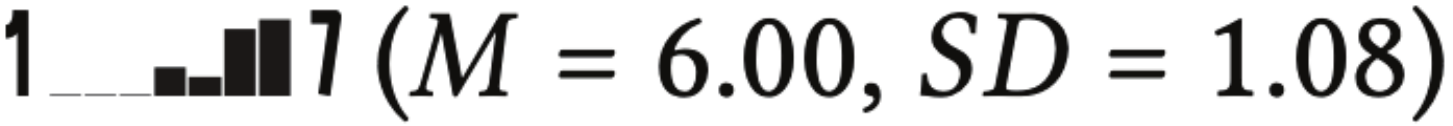}) \citePE{Shi.2021.CMA}.

Special Cases: Authors integrated graphical symbols or interface components as in-line elements within the text to establish a stronger connection between textual descriptions and figures. For example, using a \icon{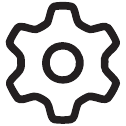} settings symbol directly in the text helps to reference the corresponding interface elements in a figure. Although such graphical elements may be originally designed to enhance interface components rather than the textual descriptions, we considered them with the function \dimension{Augmentation} if they augmented the associated text.

Among the surveyed papers, a total of 3,317 embellishments within 177 papers were found to support \dimension{Augmentation}.

\subsubsection{Text Replacement}
Paper embellishments were used to replace lengthy text as standalone elements due to their brevity. For example, when reporting the results of evaluating four timeline shapes--horizontal line~\icon{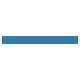}, vertical line~\icon{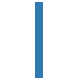}, circle~\icon{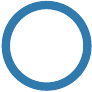}, and spiral~\icon{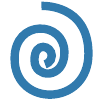}--in terms of completion time, authors used \icon{icons/circle.pdf}$<$\icon{icons/spiral.pdf} to indicate that the mean completion time for circles was significantly shorter than for spiral timelines~\cite{Bartolomeo.2020.EET}. In this case, these graphics aimed to replace condition words.

Special Cases: Many abbreviations were used to replace full names (\eg, \colorFont{D1} for a first design guideline~\citePE{Kim.2022.CDG} and \colorFont{E1} for a first expert~\citePE{Narechania.2023.DPU}). However, in these cases, the text was replaced by abbreviation rather than paper embellishments, \ie, color. Thus, we did not consider \dimension{Text Replacement} as a function of these embellishments.
Through the review, we identified 9 papers that used 414 embellishments for the \dimension{Text Replacement} function.

\subsubsection{Others}
\dimension{Others} include paper embellishments whose functions did not fit into the previous four categories. Yet, these embellishments still functioned as providing Emphasis or De-emphasis and improving Aesthetics.
For instance, authors used the same background color to mark participants' statements (\autoref{fig:examples}~\icon{icons/d.pdf}~\cite{Newburger.2024.VAS}). Similarly, authors employed the symbol \icon{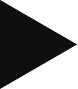} to highlight key elements from others to mark noteworthy points for future exploration~\cite{Tennekes.2021.DSO}.
Our analysis revealed 771 paper embellishments within 47 visualization papers serving for other functions.

\vspace{1em}
\textbf{Multi-functions:} While we defined four distinct primary functions of paper embellishments, they often served multiple functions simultaneously, with the most common combination being \dimension{Referencing and Differentiation}, observed in 196 papers. This is unsurprising, as embellishments that served for \dimension{Differentiation} could also facilitate \dimension{Referencing} by linking different sections of the paper or connecting text with the summary figures. Beyond this primary combination, 112 papers also incorporated \dimension{Augmentation} alongside \dimension{Referencing and Differentiation}, where embellishments typically introduced additional information or examples to reinforce distinctions between each other. Other frequently observed combinations include \dimension{Referencing and Augmentation} (42 papers) and \dimension{Differentiation and Augmentation} (29 papers). Typical examples of these function combinations are the embellishments in \autoref{fig:examples} \icon{icons/i.pdf} and \icon{icons/g.pdf}. In the first case, authors used embellishments to \dimension{Augment} the word indicating functionality and \dimension{Reference} the interface figure~\cite{Li.2023.NND}. In the second case, embellishments \dimension{Differentiated} and \dimension{Augmented} dimensions~\cite{Dhanoa.2022.PMD}.
\dimension{Text Replacement} is less commonly combined, appearing with
other functions in only 8 papers, primarily in the context of replacing text with word-scale graphics.

\subsection{\how: Design Type of Paper Embellishments}
\label{sec:HOW}
We classified the types of paper embellishments based on how they were applied---whether they were integrated into the text as additional visual elements, \eg, an icon \icon{icons/customized_icon.pdf} placed between words, or applied to text as a formatting style, \eg, altering \colorFont{text color} or adding a \background{background} highlight.
We considered inline-height embellishments that are embedded directly within the text rather than applied to it as \dimension{Word-scale Graphics}~\cite{Goffin:2017:ESW}. Meanwhile, embellishments that were applied to the text to change their appearance were classified as \dimension{Styled Words} and \dimension{Styled Paragraphs}, respectively.

\subsubsection{\wordScaleGraphics\ Word-scale Graphics}
We divided Word-scale Graphics embellishments into two categories based on whether they conveyed data or not: \dimension{Data-driven Visualization} and \dimension{Non-data-driven Graphics}, following the categorization by Goffin et al.~\cite{Goffin:2017:ESW}. We found 4,420 Word-scale Graphics embellishments in 228 papers ($\sim$\minipie{61}{black!50}{\small 61\%}).

\paragraph{\dataDriven\ Data-driven Visualizations}
encode actual data. Actual data refers to raw information that authors generated or collected from studies or observations. Thus, these embellishments typically represented and encoded paper-specific data in a compact form. The word-scale pie chart we used above\minipie{61}{black!50}{\small 61\%} is a typical example.
Likewise, researchers employed miniature charts \icon{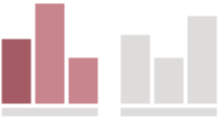} referring to a larger and more detailed chart~\citePE{Xiong.2022.VAB}, and used word-scale visualization to display additional information, \eg, adding the distribution data \icon{icons/data_driven_ex.pdf} \cite{Shi.2021.CMA}. 

Spcial Cases: One paper used filled black circles to indicate levels of complexity~\cite{Blumenschein.2020.VDH}, where \icon{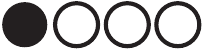} represented the lowest complexity, while \icon{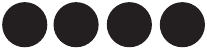} denoted the highest complexity. The visualization encoded categorical data for analysis tasks, so we categorized them in \dimension{Data-driven Visualization}.
In another example authors~\citePE{Saffo.2024.UDS} added background colors to numbers to augment this number, as shown in \icon{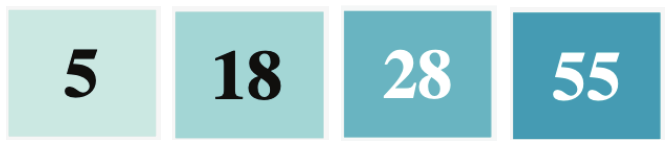}(5,18,28,55 in this illustration). While this approach changed the appearance of text, the use of color as a data-encoding mechanism led us to categorize it as \dimension{Data-driven Visualization}.
We identified 244 data-driven visualizations in 15 papers.

\paragraph{\nonDataDriven\ Non-data-driven Graphics}
are visual components that do not encode actual data themselves. This subcode spans a range of symbols, from colored blocks \tikz \fill[styledColor] (0,0) rectangle (0.2,0.2); and simple geometric symbols~\inlinevis{-1pt}{1em}{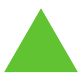} to icons of system operation buttons~\inlinevis{-1pt}{1em}{icons/settings.pdf} and pictorial representations~\inlinevis{-1pt}{1em}{icons/customized_icon.pdf}.

Special Cases: The first special case is a colormap~\icon{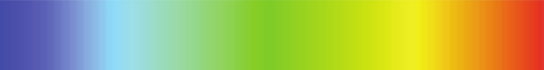} \cite{Zeng.2023.RCG} which did not encode data but served as a representation of a condition in  empirical studies. In another case authors utilized linear shapes to demonstrate data values and trends \inlinevis{-1.5pt}{1em}{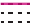}~\inlinevis{-1.5pt}{1em}{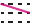}~\inlinevis{-1.5pt}{1em}{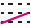}~\citePE{Sperrle.2021.LCU}. Inspired by Chen et al.~\cite{Chen:2022:NEY} we categorized such graphics as schematic representations and included as \dimension{Non-data-driven Graphics}.

4,176 embellishments in 213 papers are \dimension{Non-data-driven Graphics}, accounting for $\sim$\minipie{94}{black!50}{\small 94\%} of all \dimension{Word-scale Graphics}.

\subsubsection{\styledWords\ Styled Words}
When paper embellishments applied visual enhancements on specific text segments within a paragraph, affecting only selected words or phrases rather than the entire paragraph, we referred to them as \dimension{Styled Words}. Based on the styling type, we further classified \dimension{Styled Words} into \colorFont{Colored Fonts}, \background{Background Colors}, and \outline{Colored Outlines}.

We observed 9,421 \dimension{Styled Words} within 336 papers ($\sim$\minipie{90}{black!50}{\small 90\%}). Specifically, these embellishments include 5,503 \dimension{Colored Fonts}, 2,409 \dimension{Background Colors}, and 1,509 \dimension{Colored Outlines}.

\paragraph{\colorFont{Colored Fonts}}
is one of the most common embellishments applied to words, where authors changed the text color. We found \colorFont{Colored Fonts} in 194 papers.

\paragraph{\background{Background Colors}}
refers to the method that adds a color, other than white, behind the words. We observed texts with \background{Background Colors} in 98 papers.

\paragraph{\outline{Colored Outlines}}
refer to colored outlines around text segments, which is also classified as a type of \dimension{Styled Words}. They could take the form of a \textcolor{styledColor}{\rule{2pt}{0.8em}} left, right \textcolor{styledColor}{\rule{2pt}{0.8em}} , or \textcolor{styledColor}{\underline{\textcolor{black}{colored underline}}}, or a combination of these boundaries.
We saw 44 papers used \outline{Colored Outlines}.

\subsubsection{\styledPara\ Styled Paragraphs}
\colorbox{styledColor!10}{\parbox{\linewidth}{When the embellishments are applied to entire paragraphs, which are independent units with a new line, indentation, or numbering, we refer to them as \styledPara\ Styled Paragraphs, as seen in this paragraph.}}

A \dimension{Styled Paragraphs} incorporates paragraph-wide embellishments, such as \dimension{Colored Fonts}, \dimension{Background Colors}, \dimension{Colored Outlines}. Because the embellishments in \dimension{Styled Paragraphs} often work in combination, as opposed to isolated elements, their primary function is to enhance the paragraph as a whole instead of individual components, we did not further code the types of embellishments used in \dimension{Styled Paragraphs}.
In total, we found 191 \dimension{Styled Paragraphs} across 26 papers in this survey.

\vspace{1em}
\textbf{Multi-designs:} Authors also combined multiple types of embellishment design types, \eg, a word-scale graphic inserted alongside a styled word as shown in \autoref{fig:examples}~\icon{icons/g.pdf}. It is likely that authors intended to enhance the visual saliency of the text for greater emphasis. For instance, 47 papers combined \dimension{Background Colors} and \dimension{Colored Fonts}. A typical example is styled code snippets within text \inlinevis{-1.5pt}{1em}{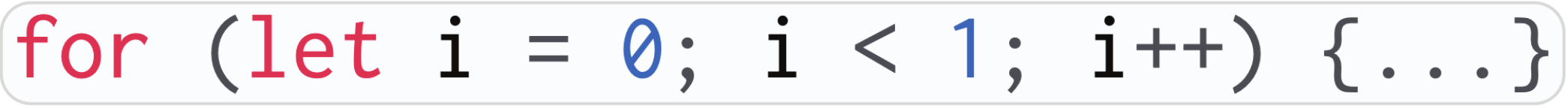}~\citePE{Hayatpur.2023.CCM}. We also observed the combination of \dimension{Word-scale Graphics} and \dimension{Colored Fonts} in 14 papers, such as symbols referring to system buttons \textcolor{orange}{new}~\inlinevis{-1.5pt}{1em}{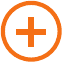}~\citePE{Wang.2024.DFA} and indicators \inlinevis{-1.5pt}{1em}{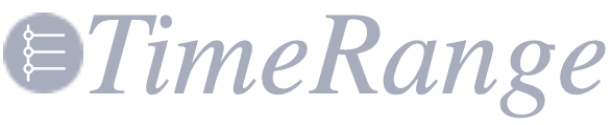}~\citePE{Zhang.2023.CVA}. 
We systematically multi-tagged design types of paper embellishments to capture their combination pattern and understand the served functions.

\subsection{\where: Location of Paper Embellishments}
\label{sec:WHERE}
We further examined where authors placed embellishments in their papers. Initially, we attempted to categorize embellishments by specific sections, such as methodology, approach, and evaluation. However, we quickly realized that visualization papers do not adhere to a uniform structure, making it challenging to assign embellishments to a fixed set of sections. For instance, in empirical studies, a ``study'' section is considered part of the methodology and approach, whereas in application papers, the same section functions as a validation. Given these structural variations across different contribution types, we took a more adaptive approach to analyzing \where embellishments appear. Specifically, we examined two key aspects: (1) the primary type of contribution the paper made and (2) the specific locations in which embellishments are placed, with a simplified version, categorized into \dimension{Structural Elements} and \dimension{Paper Sections}. This dual perspective allows us to capture how embellishments are distributed within papers of different contribution types and how their placement aligns with the overall structure of scientific writing in visualization research.

\subsubsection{Contribution Types}
We classified all papers into five categories based on each paper's primary contributions,
considering both the authors' stated contributions and the relative proportion, \ie, content length, of each contribution within the paper.
In general, a paper is assigned with a single contribution type. However, we also observed that some papers contributed almost equally to more than one type, where both contributions were highlighted by the authors, and the content was distributed roughly evenly across pages.
We multi-tagged these papers' contribution types during our coding process.
Among our collected visualization papers, we identified 143 papers contributed with \dimension{Technique}, 120 papers contributed to \dimension{Application}, 99 papers for \dimension{Empirical Study}, 43 for \dimension{Literature Review}, and 11 for \dimension{Others}. We counted papers according to their contributions because one paper can have multiple contributions.

\vspace{-0.2em}
\paragraph{Technique} papers presenting new methods or approaches for a problem or significantly extending existing techniques. Methods were usually concrete ways to solve particular problems, such as algorithms, visualization techniques, and interaction techniques. Approaches covered broader strategies for thinking about or solving a set of related problems, \eg, frameworks, workflows, and pipelines.

\vspace{-0.2em}
\paragraph{Application} papers contributed systems, applications, or interfaces. These papers typically demonstrated how visualization methods are applied to solving real-world challenges.

\vspace{-0.2em}
\paragraph{Empirical Study} papers involved collecting and analyzing data from real-world observations or experiments to answer a research question or test a hypothesis. 
This subcode also includes case studies, comparative analysis, and surveys that gathered data with questionnaires, interviews, or observations.
\vspace{-0.2em}
\paragraph{Literature Review} papers offered a systematic review, analysis, and documentation of existing visualization research. They did not collect new real-world data but instead focused on reviewing published papers and established knowledge. Their outcomes can include taxonomy, design space, or research agendas.
\vspace{-0.2em}
\paragraph{Others} included papers that make contributions not covered by the four aforementioned categories, including resources to support visualization research, such as benchmarks~\cite{Chen.2025.VEB} and datasets~\cite{Zhang.2024.OVO}.

\subsubsection{Detailed Locations}
We observed two main subcodes of locations where paper embellishments were placed: \dimension{Paper Sections} and \dimension{Structural Elements}.

\dimension{Paper Sections} refer to the main content of a paper, including Abstract, Introduction, Related Work, Core Content, and Discussion \& Conclusion. For visualization papers that did not strictly follow this structure, we applied our best judgment to align their content with the corresponding sections in our coding.
Most paper embellishments were in \dimension{Paper Sections}, with a total of 10,547 embellishments across 333 papers, representing approximately $\sim$\minipie{87}{black!50}{\small 87\%} of all used paper embellishments.
Within the \dimension{Paper Sections}, the majority of embellishments were in the Core Content, accounting for about \minipie{88}{black!50}{\small 88\%}.

\dimension{Structural Elements} are meta-textual components that define the structure and flow of a paper. They include paper titles, section headings, and figure and table captions. We documented 1,613 embellishments from 183 papers used in \dimension{Structural Elements}.

\subsection{Analysis}

\begin{figure*}[t]
    \centering
    \includegraphics[width=\linewidth]{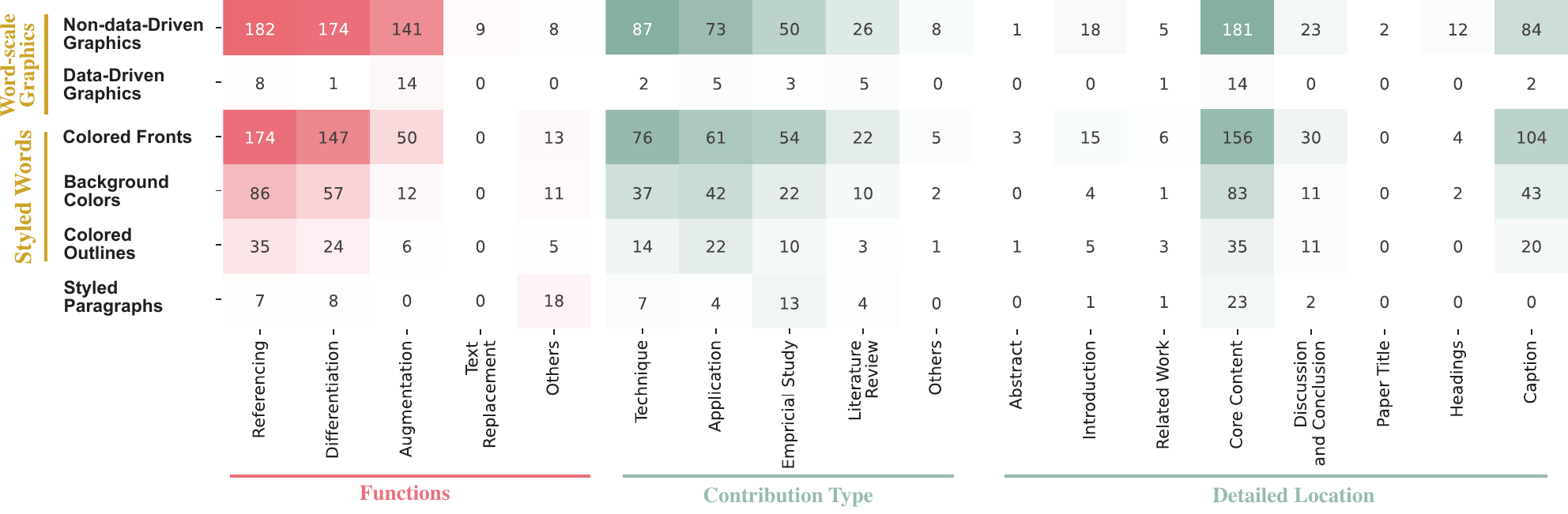}
    \caption{A heatmap illustrating the distribution of visualization papers using different types of embellishments (\how) for diverse functions (\why) and in varied contribution types of paper and locations (\where). The intensity of the color in each cell corresponds to the counts of papers, with higher values indicated by less transparency.}
    \label{fig:what_where_how}
    \vspace{-1em}
\end{figure*}

Our review of the collection revealed that the design of embellishments is influenced by both their function and placement. To better understand their usage, we conducted a cross-analysis of the \why and \where dimensions in relation to \how (\autoref{fig:what_where_how}). Since some papers included multiple tags within a single dimension, we broke down these multi-tags to analyze individual interplay.

\paragraph{\why \& \how} We observed distinct patterns in the use of embellishments for various purposes. 
To achieve \dimension{Referencing} or \dimension{Differentiation}, the most commonly used embellishments were \dimension{Non-data-driven Graphics}, appearing in $\sim$\minipie{49}{black!50}{\small 49\%} of papers for \dimension{Referencing} and $\sim$\minipie{47}{black!50}{\small 47\%} for \dimension{Differentiation}, as well as \dimension{Colored Fonts}, found in $\sim$\minipie{47}{black!50}{\small 47\%} and $\sim$\minipie{39}{black!50}{\small 39\%} of papers, respectively.
Interestingly, these two types of embellishments normally served multiple functions simultaneously (\eg, \autoref{fig:teaser}~\icon{icons/a.pdf}\icon{icons/c.pdf}\icon{icons/f.pdf}).
Specifically, 93 papers used \dimension{Non-data-driven Graphics} for \dimension{[Reference, Differentiation, Augmentation]}, while 116 papers adopted \dimension{Colored Fonts} for \dimension{[Reference, Differentiation]}.
Additionally, authors used \dimension{Colored Outlines} mostly for the combined functions of \dimension{[Reference, Differentiation]} (16 papers), though their occurrences for this combined function were less frequent than other types of styled words and \dimension{Non-data-drvien Graphics}.
Notably, for the only purpose of \dimension{Referencing}, authors favored \dimension{Styled Words} over other embellishment types: around \minipie{83}{black!50}{\small 83\%} of paper embellishments used for referencing were \dimension{Styled Words}.
However, \dimension{Styled Paragraphs} were mostly applied for other functions, \eg, highlighting study participants' statements or research questions.

This widespread embellishing of papers with colors and graphics for \dimension{Referencing} and \dimension{Differentiation} may be due to an intention to improve clarity and communication efficiency.
\edit{Color could help highlight key information, making distinctions more visually salient~\cite{Kim.2006.SGE} and potentially enabling readers to quickly identify associations and locate relevant content within extensive text~\cite{Sultanum.2025.FII}.}
Similarly, graphics \edit{might} provide an intuitive way to distinguish between different textual components, offering instant recognition. \edit{It is also possible that} well-designed \nonDataDriven\ Non-data-driven Graphics not only reinforce meaning but \edit{may contribute to improved} comprehension and memorization (serve for \dimension{[Augmentation]}).
Additionally, \dimension{Non-data-driven Graphics} \edit{appear particularly suited for} \dimension{[Text Replacement]} due to their lightweight nature. Their minimal visual presence allows them to save space by substituting lengthy text.

\paragraph{\how \& \where} We examined the usage of diverse types of paper embellishments across different parts of a paper and various contribution types. Our findings show that \dimension{Structural Elements}, which generally consist of smaller text sections than paper main body, predominantly use \dimension{Colored Fonts} ($\sim$\minipie{40}{black!50}{\small 40\%}), followed by \dimension{Non-data-driven Graphics} ($\sim$\minipie{35}{black!50}{\small 35\%}). Conversely, for \dimension{Paper Sections}, which typically involve longer and more detailed text,
these two design types of embellishments reverse the prevalence. \dimension{Non-data-driven Graphics} are most frequently used ($\sim$\minipie{36}{black!50}{\small 36\%}), with \dimension{Colored Fonts} coming in second ($\sim$\minipie{33}{black!50}{\small 33\%}). 
It may be because \dimension{Structural Elements} (\ie, headings and captions) were more visually distinct from the dense main body text, allowing embellished content in there to stand out more easily. This enhanced visibility \edit{could likely improve} recognition, making \edit{texts} more effective for functions like differentiation and referencing.
Yet, for \dimension{Paper Sections}, \dimension{Non-data-driven Graphics} may \edit{offer certain advantages over \dimension{Colored Fonts} in serving as visual anchors. As visuals could convey richer information, they might help express key points while maintaining the reading flow~\cite{Goffin:2017:ESW}.}
Further analysis of embellishment placement within a paper reveals predictable patterns: \dimension{Core Content}, the largest portion of \dimension{Paper Sections}, primarily employed \dimension{Non-data-driven Graphics}, followed by \dimension{Colored Fonts}. Similarly, \dimension{Captions}, the most extensive part of \dimension{Structural Elements}, was the second most embellished section, with \dimension{Colored Fonts} being the most frequently applied.

Across different contribution types, papers focused on \dimension{Technique} incorporated the most embellishments, with \dimension{Non-data-driven Graphics} being the most frequently used ($\sim$\minipie{39}{black!50}{\small 39\%}), followed by \dimension{Colored Fonts} ($\sim$\minipie{34}{black!50}{\small 34\%}). Likewise, \dimension{Application} papers also rely heavily on \dimension{Non-data-driven Graphics} ($\sim$\minipie{35}{black!50}{\small 35\%}) for embellishing, while also making substantial use of \dimension{Colored Fonts} ($\sim$\minipie{29}{black!50}{\small 29\%}) and \dimension{Background Colors} ($\sim$\minipie{20}{black!50}{\small 20\%}).
Additionally, \dimension{Styled Paragraphs} were most commonly used in \dimension{Empirical Studies}, as these papers often contained extensive records of experimental observations. \dimension{Styled paragraphs} helped differentiate these records from the main body content, such as methodology, design, and results, ensuring clearer organization and readability.
Among all embellishment types, \dimension{Data-driven Visualizations} were the least frequently used. This was likely because they were generally utilized for presenting less essential statistical data. In \textit{Application} and \textit{Techniques} papers, statistical reporting is relatively minimal. Meanwhile, in \dimension{Empirical Study} and \dimension{Literature Review} papers, important statistical data was often presented as full-size figures or tables, as they are closely tied to the paper's key findings and require greater emphasis.

\paragraph{\why \& \how \& \where} Despite the widespread use of various paper embellishment designs across different contribution types of visualization papers, certain patterns have emerged. One of the most common involves \dimension{Technique} papers utilizing \dimension{Non-data-driven Graphics} for the \dimension{Referencing} and \dimension{Differentiation} functions, as seen in 73 and 71 papers respectively. Similarly, \dimension{Application} papers also frequently employed \dimension{Non-data-driven Graphics} for these two functions (shown in 69 and 64 papers). This suggests that in technique- and application-oriented research, graphical elements helped link textual descriptions to corresponding objects, such as interface elements~\cite{Chen.2023.BAB} or interactive buttons~\cite{Li.2023.NND}.
They could help clarify the design or functionality of techniques or applications, \edit{potentially supporting} users' understanding and memory retention of them \edit{as we mentioned in \autoref{sec: opportunities_challenges}}.
However, instead of \dimension{Non-data-driven Graphics}, \dimension{Empirical Studies} majorly used \dimension{Colored Fonts} for \dimension{Referencing} (48 papers) and \dimension{Differentiation} (46 papers). This is understandable, as \dimension{Empirical Studies} often involve detailed experimental conditions and nuanced findings that require clear distinctions. \dimension{Colored Fonts} \edit{potentially supported} quick identification and differentiation of key elements within the text, making them\redsout{particularly} effective for referencing related ideas and promoting clarity without disrupting the flow of the content.
Moreover, \dimension{Literature Review} papers showed a preference for using \dimension{Non-data-driven Graphics}, particularly for \dimension{Differentiation} (22 papers), \dimension{Referencing} (21 papers), and \dimension{Augmentation} (18 papers). Given that \dimension{Literature Reviews} usually proposed taxonomies, these embellishments likely served to visually reinforce categorical distinctions. \dimension{Non-data-driven Graphics} appeared effective in illustrating these classifications, enhancing the presentation of taxonomic structures.

\section{\edit{Opportunities and Challenges}}
\label{sec: opportunities_challenges}
\edit{Having outlined the current design practices of paper embellishments, we now turn to a broader reflection on their implications. Since few studies have directly examined the impact of paper embellishments on scientific reading, we focus on research that investigates visual embellishments in other contexts, such as charts and general written content. By synthesizing insights from these studies along with our observations of current practices in visualization papers, we identify a set of potential opportunities and challenges associated with the use of paper embellishments in scientific writing.}

\subsection{Opportunities}
\edit{\textbf{Better Text-data Linking.} The interview findings by Sultanum et al.~\cite{Sultanum.2025.FII} revealed that designers often use aligned text formatting alongside visual marks to create stronger connections between textual and visual elements within dashboards. Similarly, Latif and Beck~\cite{Latif:2019:VAD} found that placing large-scale images away from the relevant text can disrupt the reading flow, causing a split-attention effect and an increase in cognitive burden. These findings suggest that incorporating styled text or integrated embellishments directly within scientific articles could help readers more effectively connect textual descriptions with the corresponding data.} 

\noindent\edit{\textbf{Better Recall.} If paper embellishments could improve text-data linking, as suggested earlier, they may also support better recall. Zhi et al.~\cite{Zhi.2019.LLE} found that enhanced lineage between data and text improves readers' ability to remember information. Similarly, Bateman et al.~\cite{Bateman:2010:UJE} demonstrated that embellished charts led to improved recall performance compared to plain ones. These findings suggest that, when applied to scientific publications, paper embellishments could help readers retain key information more effectively.}

\noindent\edit{\textbf{Enhancing Readability.} Kobayashi and Kawashima~\cite{Kobayashi.2019.PFT} investigated the effects of sequentially fading out words during reading and found that this technique encouraged readers to focus on the first sentence of each paragraph, similar to using colored fonts, potentially leading to improved comprehension. Building on this idea, Joshi and Vogel~\cite{Joshi.2024.CHD} examined the effects of allowing readers to interact with textual embellishments, \ie, textual background highlighting. Their results showed that limiting the number of highlighted words in a document helped readers better retain key information. While these studies focus on textual styling rather than in-line visualizations, various forms of paper embellishments may enhance readability more broadly.}

\noindent\edit{\textbf{Cognitive Ease.} In programming environments, visual enhancements, such as syntax highlighting through colors, text styling, and graphics, have been shown to improve task performance without introducing visual overload~\cite{Sulir:2018:VAS, Asenov.2016.ERV}. Similarly, in the domain of language learning, textual enhancements such as underlining, boldface, and italics have been found to support comprehension and learning in second-language contexts without hindering understanding~\cite{Labrozzi:2016:ETE, Meguro:2019:TEG}. These findings suggest that, in scientific writing, such visual embellishments could enhance the reading experience while minimizing additional cognitive load. In addition, when figures and descriptive texts are spatially separated, readers often experience a split-attention effect, which increases cognitive burden~\cite{Beck.2017.WSG}. Integrating word-sized graphics directly into the text has been shown to reduce this effect and promote cognitive ease~\cite{Ayres:2005:SAP, Ginns:2006:IIM}. Moreover, Barach et al.~\cite{Barach.2021.EPW} found that readers processed target words more efficiently when emojis provided helpful context or disambiguation, evidenced by shorter fixations and more frequent skipping. Since emojis embedded in text also constitute a form of paper embellishment, this further supports their potential to facilitate smoother reading and reduce mental effort.}

\subsection{Challenges}
\edit{\textbf{Distraction.} Tufte~\cite{Tufte.1983.VDQ} criticized visual clutter, \eg, moiré vibration, for introducing uncontrolled ambiguity. These forms of ``chart junk'' fail to enhance the effectiveness of visual communication and can even degrade the clarity of data presentation. Such distractions may appear in graphics, \eg, improper cross-hatching, or in textual elements, \eg, jagged all-uppercase sans-serif lettering. Similarly, Goffin et al.~\cite{Goffin.2014.EPD} found that in-line visualizations may interrupt sentence flow and make the surrounding text more difficult to interpret in some cases. Building on these observations, we see a potential challenge in integrating visuals within scientific texts: when poorly designed, paper embellishments can distract readers, offer little to no informational benefit, and ultimately detract from the message being conveyed.}

\noindent\edit{\textbf{Visually Overstimulating.} The study exploring the effects of embellishments in charts for non-experts found that the saliency of embellishments influenced both perceived appeal and ease of understanding~\cite{Alebri:2024:ERP}. Applying their definition, paper embellishments can likewise vary in visual prominence within a scientific paper. When these elements exhibit high saliency, they may draw disproportionate attention, potentially leading to visual overstimulation and cognitive fatigue. This highlights the need to strike a careful balance where embellishments should enhance clarity and presentation without overwhelming the reader or distracting from the core content.}

\noindent\edit{\textbf{Static Media.} As found by Head et al.~\cite{Head:2022:MAH}, when authors design custom mathematical augmentations, they often aim to show multiple levels of information. However, static media inherently limit reader interaction, restricting the gradual exploration or unfolding of information over time. This constraint poses a challenge for paper embellishment, as while they may convey additional layers of meaning, the static nature of printed or non-interactive formats can hinder readers from engaging with them dynamically, potentially reducing their effectiveness.}

\section{Discussion}
\redsout{To provide future authors with \edit{inspirations}\redsout{practical examples for choosing paper embellishments to enhance scientific communication and improve paper readability}, we \edit{subjectively selected} representative examples illustrating the application of our \edit{characterizations}. We \redsout{first} chose interesting and promising examples collected from our literature review. \redsout{Then, we further analyzed paper embellishments that were used in our own previously published or submitted papers, allowing us to explain the authors' perspective in describing the design choices and including reviewer feedback on these embellishments.} \edit{We present a variety of cases in \autoref{fig:examples}} that we identified as typical and representative paper embellishments in our collections.}
By examining the current strategies authors used to design and place paper embellishments in their publications,
\edit{we identify and discuss four common design patterns among all the examples.}
\redsout{design strategies we find useful to consider and explore further.}

\begin{itemize}
    \item \edit{\textbf{The use of actual colors instead of referring to them by name.} We saw many instances of embedding color blocks alongside their corresponding names or applying colors directly to the text. The colors usually appear both in the figures and textual descriptions. As noted by Berlin and Kay~\cite{Berlin.1991.BCT}, individuals from different cultural backgrounds may perceive and describe colors differently, making precise color descriptions challenging. When referring to multiple colors within a figure, relying solely on textual descriptions can introduce ambiguity and misinterpretation. This challenge may become even more pronounced when authors use a large number of colors, such as 10 or more~\citePE{Chatzimparmpas.2020.TSA}, which are frequently referenced in the text. Though not tested, we see significant potential in enhancing color communication by displaying colors. Incorporating small colored dots or blocks adjacent to the text, or applying color directly to the terms, could probably provide an intuitive and immediate visual reference. Thus, they could potentially not only enhance clarity but also strengthen the connection between text and figures, ultimately improving readability and comprehension.}
    \item \edit{\textbf{The use of consistent color themes for strongly related textual components.} A lot of papers used a uniform color theme~\citePE{Kim.2021.GGR, Dhanoa.2022.PMD}. It} can potentially help readers mentally connect related concepts more effectively \edit{based on the findings from previous work~\cite{Qu.2018.KMV}}.
    However, using similar or even the same color for loosely related elements may increase cognitive load, making it more difficult for readers to distinguish them. For example, if key terms are colored similarly to default hyperlink colors~\cite{Fygenson.2024.AMI}, readers might mistakenly interpret them as clickable links rather than emphasized content. 
    \item \edit{\textbf{The use of Word-scale Graphics for immediate visual cues.}} \edit{As we mentioned in \autoref{sec:HOW}, word-scale graphics is the most popular design type of paper embellishments. It is understandable because using easily recognizable imagery icons could enhance learning and comprehension, as suggested by the study~\cite{Chajadi.2020.EVD}.} While word-scale graphics in scientific papers are usually carefully designed, \edit{we found that} their size constraints may make finer details difficult to discern, potentially hindering differentiation between visual elements. To \edit{avoid this problem, we saw some papers designing word-scale graphics which are} immediately recognizable and easily interpretable~\citePE{Dhanoa.2022.PMD}. For more complex visuals, minifigures might be a good replacement to enhance readability.
    \redsout{Additionally, it is important to design graphics with multiple formats in mind, such as print, digital, HTML, and PDF, as each medium may render them differently in terms of scale and resolution.}
    \item \edit{\textbf{Visual hierarchy and consistency of paper embellishments.} We noticed that authors normally keep consistency in color or texture across distinct but related elements, probably aiming to convey categorization information through the design itself~\citePE{Saffo.2024.UDS, Zhang.2023.CVA}. It would potentially enhance the connection between related elements within the same category to share a similar design style. Likewise, using embellishments to emphasize key content could help establish a clear visual hierarchy.} \edit{Schulz et al.~\cite{Schulz.2011.DSI} reviewed hierarchical data visualization techniques and highlighted that ineffective hierarchy design can obscure relationships and hinder comprehension. This insight is equally relevant to paper embellishments, as textual content also follows an inherent hierarchical structure.} Some information may be more critical than others, and certain categories may serve as overarching groups for subcategories. However, despite the widespread use of embellishments, few are explicitly designed to convey hierarchical relationships. A potential approach could be to adopt a consistent general design for elements within the same category and then apply the same style to elements at the same hierarchical level while differentiating styles for elements at different levels. For example, adjusting the saliency or brightness of a hue can visually signify different levels within a hierarchy, providing subtle yet effective cues for structuring information~\citePE{Saffo.2024.UDS}. Alternatively, \edit{another approach we found in the collection is placing different numbers} of embellishments before the corresponding text to indicate the hierarchical level~\citePE{Hong.2024.SDC}. The authors designed visual representations for sub-dimensions, while top-level dimensions did not include graphics. For sub-sub-dimensions, they used two icons, with one representing the corresponding sub-dimension and the other representing the sub-sub-dimension. \redsout{They also applied a consistent color scheme to the icons within the same top-level dimension for visual coherence.}
\end{itemize}

\redsout{In this section, we highlight \edit{interesting} findings\redsout{and design suggestions} that we observed from the review process.}

\edit{Beyond design patterns, we found that both authors and publishers play important roles in effectively leveraging paper embellishments. We \textbf{encourage publishers} to provide updated authoring resources and to ensure the proper and consistent rendering of embellishments across digital platforms.} While PDF versions generally preserve line-height symbols and icons within texts as intended, web-based HTML previews often introduce inconsistencies. \edit{For instance, on IEEE's web preview, inline symbols and icons appear significantly larger, usually occupying more than one line, which results in inconsistent line spacing and disrupts the reading flow~\citePE{Vieth.2024.MVF, Scimone.2024.MVT}}. In contrast, ACM CHI and CGF's HTML versions maintain properly scaled, inline symbols, ensuring a more seamless reading experience~\citePE{Shi.2021.CMA, Liebers.2023.VCM}. Standardizing how embellishments appear in both digital previews and downloadable versions would help maintain their intended \edit{designed} function.\redsout{ and, therefore, ensure they enhance comprehension rather than inadvertently hindering it}

\edit{Beyond static presentation, \textbf{interaction with paper embellishments} could play a significant role~\cite{Beck.2017.WSG}, especially on emerging media platforms~\cite{Hohman.2020.CIA}. Prior work has shown that interactive tutorials can lead to faster reading and fewer errors~\cite{Kelleher.2005.SBT}. In the scientific publication field, Distill~\cite{Distill} pioneered the integration of interactive visualizations directly into papers, allowing readers to engage with the algorithms being reported. Although Distill ceased publication in 2021 due to the substantial human effort required for curating such content, it remains a compelling example of how interactivity can transform complicated scientific communication.}

While incorporating embellishments into our manuscript, we discovered that \textbf{the process can be time-consuming}. The most demanding aspects include deciding whether and what embellishments to add, designing the graphics, and researching appropriate \LaTeX\ libraries and codes for implementation. This raises an important question about the trade-off between effort and benefit: Is it worthwhile to invest time in designing and embedding visual elements within the text? Our tests suggest that large language models (LLMs) can assist with \LaTeX\ implementation to achieve the intended design. However, effective use of LLMs requires highly specific and well-structured prompts, highlighting the need for a balance between automation and manual refinement.

\section{A Research Agenda for Paper Embellishments}
Our analysis revealed a growing adoption of paper embellishments in visualization papers. Among them, IEEE VIS 2024 had the highest proportion, with $\sim$\minipie{49}{black!50}{\small 49\%} of IEEE VIS 2024 publications using paper embellishments (see \autoref{app:paper_trend}). While paper embellishments are becoming more prevalent, particularly in visualization papers, there is limited information on their effective use. We aim to identify key directions for further exploration, including systematic evaluation, practical application, and broader investigation.

\subsection{Evaluations}
Through our systematic review, we identified several promising directions for future research. 
A key fundamental question is \textbf{whether paper embellishments can genuinely enhance papers' readability, comprehension, or memorability of the content}. According to the research done by Huth et al.~\cite{Huth.2024.ETT}, visualization techniques within text could potentially improve reading behaviors.
While our analysis showed that about 25\% of the surveyed conference publications incorporated paper embellishments, little research has been conducted to evaluate their actual impact on scientific communication.
Previous research found that certain design choices can significantly affect people's readability, such as fonts~\cite{Arditi.2005.AFL} and text colors~\cite{Garcia.1996.ECT}. 
These findings suggest the potential of well-designed paper embellishments to contribute to faster reading and improved comprehension.
Yet, it is important to collect dedicated empirical evidence for different types of paper embellishments. While evidence might exist for font types, colors, and background colors, much less is known about in-line graphics, for example.  These studies could provide valuable guidance for future authors about the effective usage of paper embellishments, helping them produce clearer and more accessible scientific papers for a broader audience. 

Beyond their potential benefits, another critical aspect to explore is \textbf{how to appropriately use paper embellishments while avoiding cognitive overload}. Research on working memory suggests that humans can retain limited units of information at a time~\cite{Miller.1956.MNS}. However, we observed a lot of embellishments, including \dimension{Non-data-driven Graphics} and \dimension{Styled Words}, appearing within a short paragraph~\cite{Hong.2024.SDC, Yang.2024.SPI}. Other examples include the use of multiple colors occurring frequently within a short span, such as numbers with background colors appearing 40 times within half a page~\cite{Lan.2021.UNL}. It is essential to assess how many embellishments are appropriate for a given text length and what limits should be set to prevent overwhelming the reader. For example, a specific question arises: How many colors should be used to effectively emphasize differences while ensuring memorability without causing distraction?

We observed \textbf{a potential gap between authors' intentions and readers' perceptions of certain paper embellishments} from our review process. For example, in~\citePE{Wang.2023.GAM}, the authors used a consistent color for all hyperlinks, including references, section links, code lines, and design goal links. While this uniformity may help readers easily identify hyperlinks, it could also lead to potential confusion. Given that the paper also employs six other distinct colors to explain code elements, some readers might interpret certain hyperlinks as part of the code. It raises an interesting question: Were the authors' intentions effectively conveyed to readers? It means that even if paper embellishments were shown to enhance readability, their effectiveness depends on how clearly authors convey their intended meaning, for example through explicit connections to reference words~\cite{Goffin:2017:ESW}. Thus, it is also crucial to investigate this communication gap and identify ways to enhance such communications via paper embellishments.

For the same functions, multiple types of embellishments can be used to achieve the desired effect. A critical next step is to conduct an evaluation study to compare the effectiveness of different embellishment strategies. For instance, in the case of \dimension{Referencing}, both \dimension{Non-data-driven Graphics} and \dimension{Colored Fonts} are commonly used. However, it remains unclear which type is most effective in various contexts. Understanding these differences through empirical studies would provide valuable guidance for authors in selecting the most appropriate embellishments for enhancing scientific communication.

These overarching directions branch into several specific research challenges, including \textbf{developing detailed design guidelines for different use cases}. For example, when distinguishing hierarchical dimensions, the number of levels in the hierarchy may influence the optimal embellishment approach. Understanding these nuances requires a series of studies investigating how different types of embellishments contribute to readability, comprehension, and knowledge retention. As previously mentioned, selecting appropriate colors can be particularly challenging when encoding a large number of distinct cases. Similarly, it is crucial to balance the quantity of embellishments with the length of the text to enhance comprehension while avoiding unnecessary distractions. This involves carefully managing the density of embellishments within the text to ensure they support, rather than hinder, the reader's understanding. 
By systematically exploring such aspects, researchers could follow more concrete recommendations to integrate embellishments effectively in scientific writing.

Another intriguing research direction is \textbf{examining whether Non-data-driven graphics, such as icons, are both sufficiently distinguishable from one another and effectively aligned with the concepts they represent}.
From our collection, we found that 51 survey papers incorporated paper embellishments when describing their dimensions. These embellishments included colored dimension labels, background-colored dimensions, and icons with or without colors. We see a strong need for incorporating icons alongside dimensions to enhance readers' understanding and improve memorization of taxonomies. However, we observed that few manuscripts have explicitly addressed the design rationale behind the icons used in scientific papers. As authors, we have also encountered reviewer feedback questioning the reasoning behind our icon design choices. This highlights a broader gap in understanding how to create icons that effectively represent keywords while remaining visually distinct from one another. Future research could explore strategies for designing icons that accurately represent concepts while remaining visually distinct from one another within paragraphs. Investigating factors such as shape, color, and semantic association could lead to more structured guidelines, helping authors make informed design choices within scientific writing.

\subsection{Applications, Libraries, and Toolkits}
To make embellishments easier to use in practice it is important to further develop applications, libraries, or toolkits that would allow to automatically suggest or even incorporate appropriate embellishments into scientific papers.
These tools could not only use customized models but also leverage the capabilities of large language models (LLMs), which have demonstrated significant potential in assisting with scientific writing~\cite{Altmae.2023.AIS, Cheng:2024:AGT}. While current LLMs can generate basic formatting elements (\eg, \textbf{bold}, \textit{italic}, emojis, $\bullet$ bullet points), they lack the ability to produce more customized and complicated paper embellishments. By integrating carefully designed prompts and interactions with LLMs into a paper authoring system, we can enhance the system to streamline the writing process for authors and enable the automatic generation of meaningful embellishments for scientific papers.
Other than automatically generating embellishments by systems, for web-based papers, embellishments could be interactive. Readers could selectively make them appear or disappear, blend in or stand out, or show details-on-demand.

Moreover, we noticed other creative ways of reusing template-provided techniques for novel applications. For example, font formatting can be used to encode data, such as varying font weights to represent numerical values, as seen in \href{http://fatfonts.org/}{FatFonts}. Similarly, different outline shapes---such as rectangles, circles, and triangles---could be leveraged to distinguish between categories~\citePE{Lohfink.2022.KRA}.
Future research can be extended to explore a wider range of embellishment techniques and assess their impact on reading experiences. By examining their usage cases, we could provide further guidance in selecting effective embellishments for scientific papers and potentially inspire the new embellishment designs.

\section{Conclusion}

In this paper, we conducted a systematic review of papers with embellishments, published in IEEE VIS, ACM CHI, and EuroVis from 2019 to 2024. By analyzing \colCount\ papers, we \edit{observed a growing trend in the use of embellishments within the visualization research community. We identified key opportunities and challenges, and} identified \edit{characterizations} structured around \why, \how, and \where. \redsout{\dimension{Non-data-driven Graphics} and \dimension{Colored Fonts} were the most commonly used paper embellishments. They were mainly designed for \dimension{Referencing} or \dimension{Differentiation}. Most embellishments appeared in \dimension{Application} or \dimension{Technique} papers within \dimension{Core Content} or \dimension{Captions} of figures or tables.} 
\edit{Importantly, our manuscript aimed to document and interpret the current landscape of paper embellishment practices, rather than to evaluate their effectiveness. As such, instead of prescribing concrete guidelines for improving the current practices, we propose a research agenda that highlights the need for future empirical studies to assess the impact of various types of embellishments on readability, comprehension, and retention. Overall, our findings lay the groundwork for a deeper understanding of the role of visual embellishments in scientific papers. Moving forward, both researchers and publishers have critical roles to play, where authors explore new ways to enhance communication through design, and publishers support flexible and innovative formats for scholarly dissemination.} \redsout{Our findings provide a foundation for understanding the role of scientific paper embellishments. Future work could further investigate their effectiveness in enhancing readability, comprehension, and retention. Beyond the contributions that authors could make to improve paper readability, publishers also play a crucial role in supporting new formats of scientific communication.} 
%









\bibliographystyle{abbrv-doi-hyperref}

\bibliography{template}
\flushcolsend
\clearpage
\appendix
\section{Exclusion examples}
\label{app:exclusion}
\renewcommand\thefigure{\thesection.\arabic{figure}} 
\setcounter{figure}{0}

\begin{figure}
    \centering
    \includegraphics[width=\linewidth]{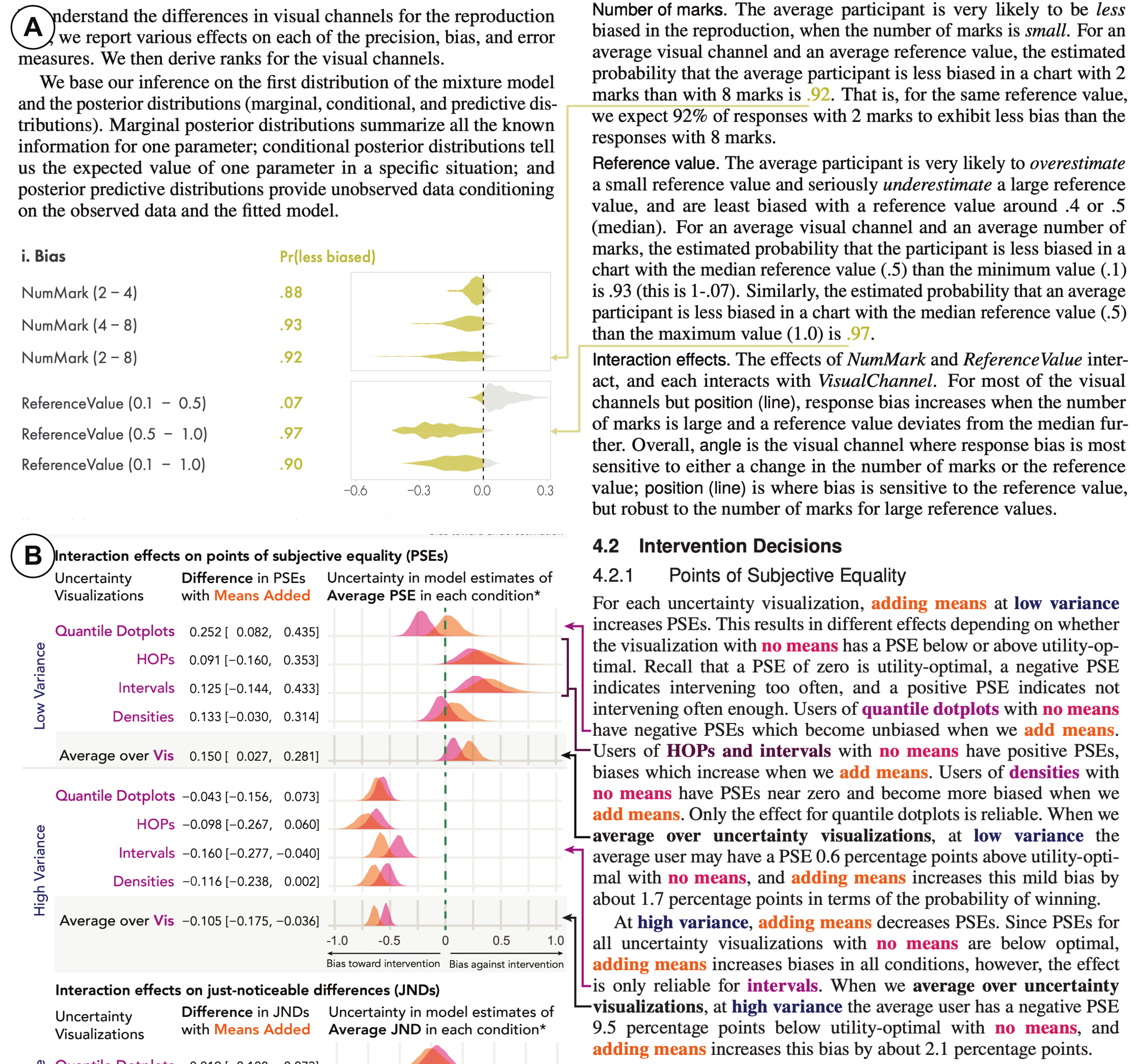}
    \caption{
    Examples for using leading lines that we excluded from our paper embellishment collections: \icon{icons/a.pdf} added leading lines between them~\cite{McColeman.2022.RRV}. \icon{icons/b.pdf} used color-coded leading lines to differentiate cases~\cite{Kale.2021.VRS}. 
    }
    \label{fig:exceptions}
\end{figure}

We present interesting examples we did not include in our collection (see \autoref{fig:exceptions}). The first two examples are leading lines between corresponding figures and descriptive texts, as shown in \autoref{fig:exceptions}~\icon{icons/a.pdf} \icon{icons/b.pdf}. 


\section{Trend of papers using embellishments}
\label{app:paper_trend}

\begin{figure}
    \centering
    \includegraphics[width=\linewidth]{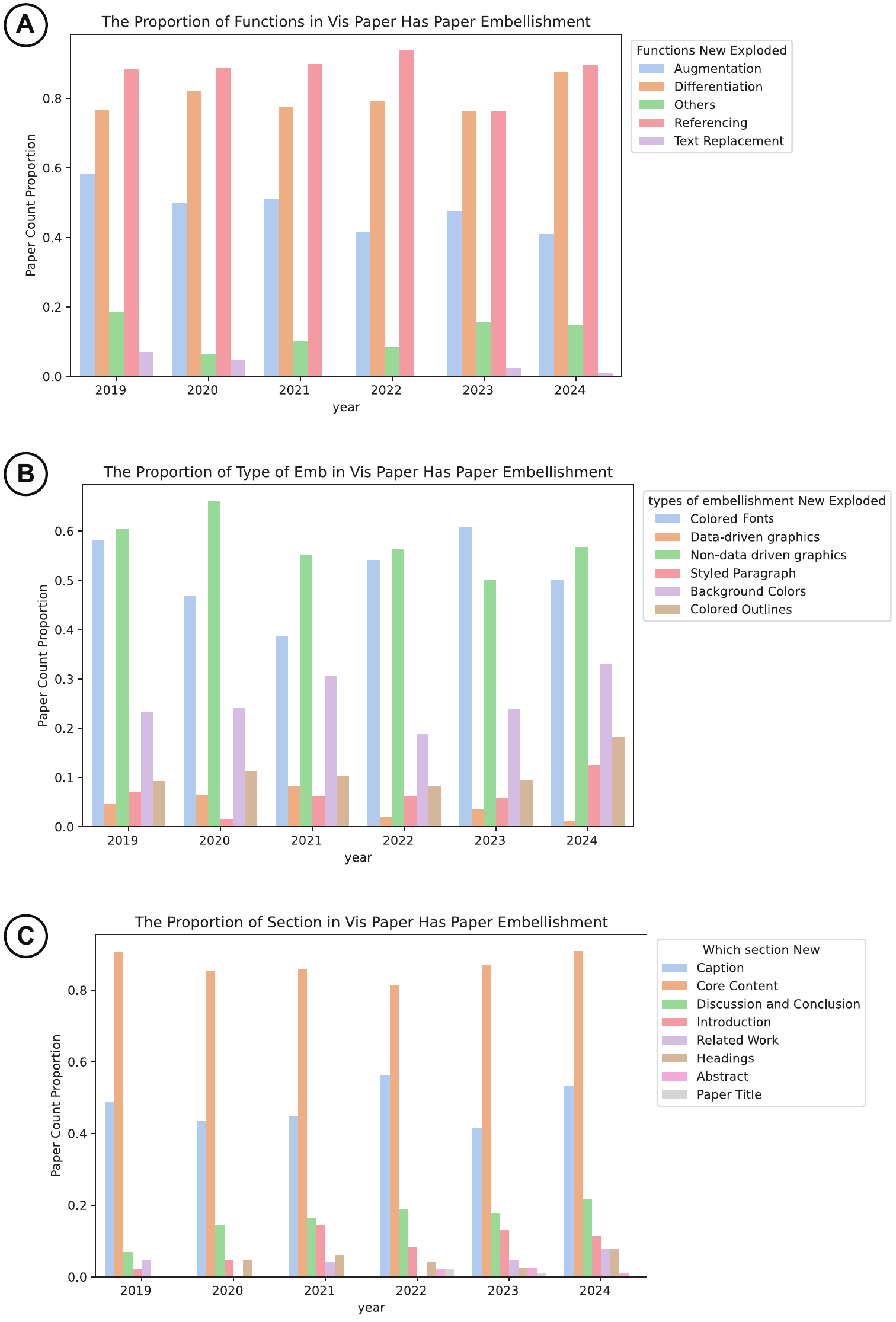}
    \caption{The trend in the proportion of publications incorporating paper embellishments. \inlinevis{-1pt}{1em}{icons/a.pdf} indicates the proportion trend of all the functions (\why) in visualization papers across six years. \icon{icons/b.pdf} shows the proportion trend of embellishment types (\how), while \icon{icons/c.pdf} represents the proportion of sections (\where).
    }
    \label{fig:app_pub_with_pe}
\end{figure}

After drilling into the different dimensions over the years (see \autoref{fig:app_pub_with_pe}), we observed that paper embellishments are mostly found in \dimension{Core Content}, followed by the \dimension{Caption} in collected papers, with little change over the past six years. This is expected, as \dimension{Core Content} forms the major body of a research paper, and \dimension{Caption} serves to better illustrate the figure or tables, helping to establish stronger connections between texts and referred visuals.

From 2019 to 2024, the proportion of paper embellishments in both the \dimension{Introduction} and \dimension{Discussion and Conclusion} sections exhibited an overall increasing trend, emphasizing the prevalence of using paper embellishment in more high-level narrative components of research papers. The proportion of embellishments in \dimension{Introduction} started low in 2019 and peaked in 2021 before experiencing slight fluctuations in the following years, settling at around 11\% in 2024.
Conversely, \dimension{Discussion and Conclusion} showed a steadier upward trend---starting at $\sim$7\% in 2019, the proportion consistently increased each year, reaching $\sim$22\% in 2024.
Surprisingly, paper embellishments were observed in paper \dimension{Abstract} and \dimension{Title} since 2022. This implies a growing trend in their use throughout the entire paper.
As for the type of paper embellishments, the usage of all types of embellishments presented fluctuating trends over six years. \dimension{Non-data-driven Graphics} and \dimension{Colored Fonts} consistently were always the most commonly used embellishments, followed by \dimension{Background Colors}.
However, in 2024, the proportion of papers using \dimension{Styled Paragraphs} and \dimension{Colored Outlines} significantly raised to their highest levels compared to previous years. In contrast, \dimension{Data-driven Graphics} declined to its lowest point.
\newpage
\bibliographystylePE{abbrv-doi-hyperref}
\bibliographyPE{collection}
\nocitePE{*}










\end{document}